\documentclass[12pt]{article}
\usepackage{graphicx}
\usepackage{amsmath}
\usepackage{cite}

\usepackage{amssymb}

\newcount\hour\newcount\minute
        \hour=\time \divide\hour by60 \minute=\time
        {\multiply\hour by60 \global\advance\minute by-\hour}
        \edef\militarytime{\number\hour:\ifnum\minute<10
0\fi\number\minute}

\makeatletter
\DeclareRobustCommand*\cal{\@fontswitch\relax\mathcal}
\makeatother

\def\draftdate{\relax}
\def\mda{\relax}
\def\mua{\relax}
\def\mla{\relax}
\def\draft{
\def\thtystars{******************************}
\def\sixtystars{\thtystars\thtystars}
\typeout{}
\typeout{\sixtystars**}
\typeout{* Draft mode!
         For final version remove \protect\draft\space in source file *}
\typeout{\sixtystars**}
\typeout{}
\def\draftdate{\today}
\def\mua{\marginpar[\boldmath\hfil$\uparrow$]%
                   {\boldmath$\uparrow$\hfil}%
                    \typeout{marginpar: $\uparrow$}\ignorespaces}
\def\mda{\marginpar[\boldmath\hfil$\downarrow$]%
                   {\boldmath$\downarrow$\hfil}%
                    \typeout{marginpar: $\downarrow$}\ignorespaces}
\def\mla{\marginpar[\boldmath\hfil$\rightarrow$]%
                   {\boldmath$\leftarrow $\hfil}%
                    \typeout{marginpar: $\leftrightarrow$}\ignorespaces}
\overfullrule 5pt
\oddsidemargin -15mm
\marginparwidth 29mm
}

\def\stars{\strut\leaders\hbox{*}\hfill\strut}
\def\starline{\hfil\strut\hfil\hbox to \textwidth {\stars}\hfil}

\newcommand\sproc      {$s$ process}

\newcommand\bdecay     {$\beta$ decay}
\newcommand\sdproc     {$s$-process}

\newcommand\Zmax       {\ensuremath{Z_{\max}}}
\newcommand\Zmaxe      {\ensuremath{Z_{\max}^{({\rm e})}}}
\newcommand\Zmaxo      {\ensuremath{Z_{\max}^{({\rm o})}}}

\newcommand\Ref[1]     {Ref.\,\cite{#1}}
\newcommand\Refs[1]    {Refs.\,\cite{#1}}
\newcommand\eqn[1]     {Eq.\,(\ref{#1})}

\newcommand\fig[1]     {Fig.\,{\ref{#1}}}

\newcommand\tab[1]     {Table~\ref{#1}}

\def\beq{\begin{equation}}
\def\eeq{\end{equation}}
\def\bsp#1\esp{\begin{split}#1\end{split}}
\def\bal#1\eal{\begin{align}#1\end{align}}
\def\beeq{\begin{eqnarray}}
\def\eeeq{\end{eqnarray}}

\begin{document}


\begin{titlepage}
\renewcommand{\thefootnote}{\fnsymbol{footnote}}
\begin{flushright}
arXiv:yymm.nnnn [nucl-ex]
\end{flushright}
\par \vspace{5mm}
\begin{center}
{\Large \bf Phenomenological description of
neutron capture cross sections at 30\,keV}
\end{center}
\par \vspace{2mm}
\begin{center}
{\bf Mikl\'os Kiss}\\[.5em]
{Berze N.J. Gimn\'azium, Kossuth 33, H-3200 Gy\"ongy\"os, Hungary\\
E-mail: kiss-m@chello.hu}
and\\[1em]
{\bf Zolt\'an Tr\'ocs\'anyi}\\[.5em]
{University of Debrecen and Institute of Nuclear Research of the
Hungarian Academy of Sciences, H-4001 Debrecen P.O.Box 51, Hungary\\
E-mail: z.trocsanyi@atomki.hu}
\vspace{3mm}
\end{center}

\par \vspace{2mm}
\begin{center} {\large \bf Abstract} \end{center}
\begin{quote}
\pretolerance 10000
Studying published data of Maxwellian averaged neutron capture
cross sections, we found simple phenomenological rules obeyed by the
cross sections as a function of proton and neutron number. We use these
rules to make predictions for cross sections of neutron capture on
nuclei with proton number above 83, where very few data are available.
\end{quote}
\vspace*{\fill}
\begin{flushleft}
     2011
\end{flushleft}
\end{titlepage}
\clearpage



\renewcommand{\thefootnote}{\fnsymbol{footnote}}


\section{Introduction}
\label{sec:intro}

Theoretical descriptions of nucleosynthesis in stars rely heavily on the
knowledge of capture cross sections of slow neutrons on nuclei. The
classical model of nucleosynthesis in weak neutron flux is based on
slow neutron capture (the \sproc) that occurs along a path in the
stability valley of nuclei (see for instance,
\Refs{Burbidge:1957vc,Cameron:1958vx,Kappeler:1989,Gallino:1998us}).
The \sdproc\ evolution codes take into account the most important
processes (those with largest cross sections) along the stability valley.
The necessary information on the neutron capture cross sections and
\bdecay\ life times, needed to describe qualitatively the abundances of
the \sdproc\ elements, is rather well known from laboratory experiments
\cite{Kappeler:2000,kadonis,Nakagawa:2005,Pritychenko:2009fe}.

The \sdproc\ model is capable to explain the observed abundance of heavy
elements fairly well \cite{Rolfs:1988}. The difference of observation and
prediction is largely attributed to another process that occurs in
stellar enviroment with high neutron flux, typically in supernovae.
In such circumstances the neutron capture is very likely and neutron
rich nuclei far from the stability valley build up very quickly due to
repeated capture of neutrons. The nuclei produced such a way are so
unstable and short-lived that experimental information about their
capture cross sections and decay life times is not generally available.

In a recent work we proposed a unified model of nucleosynthesis
of heavy elements in stars \cite{Kiss:2010}. That approach takes
into account all possible types of production and depletion mechanisms
and solves the whole system of differential equations numerically. The
result of such an approach is that (instead of the \sdproc\ path) the
evolution of the synthesis proceeds along a band in the valley of
stable nuclei. The width of this band -- and consequently the final
abundances of nuclei -- depends on the neutron flux and the
capture cross sections on individual nuclei charactherized by both their
proton and neutron numbers, $\sigma(Z,N)$, which constitutes an essential
input to the model calculations. Therefore, it is important to learn
about these cross sections as much as possible.

In this paper, we study the general features of Maxwellian averaged
neutron capture cross sections collected in recent compilations of data
\cite{Nakagawa:2005,Pritychenko:2009fe}. In section 2 we show some
phenomenological observations. In the following section we use those to
make some order of magnitude predictions for the cature cross sections
$\sigma(Z,N)$ for proton numbers $Z>83$, where only very few data are
available. Section 4 contains our conclusions.

\section{Observations}

\begin{figure}[ht!]
\includegraphics[width=.49\textwidth]{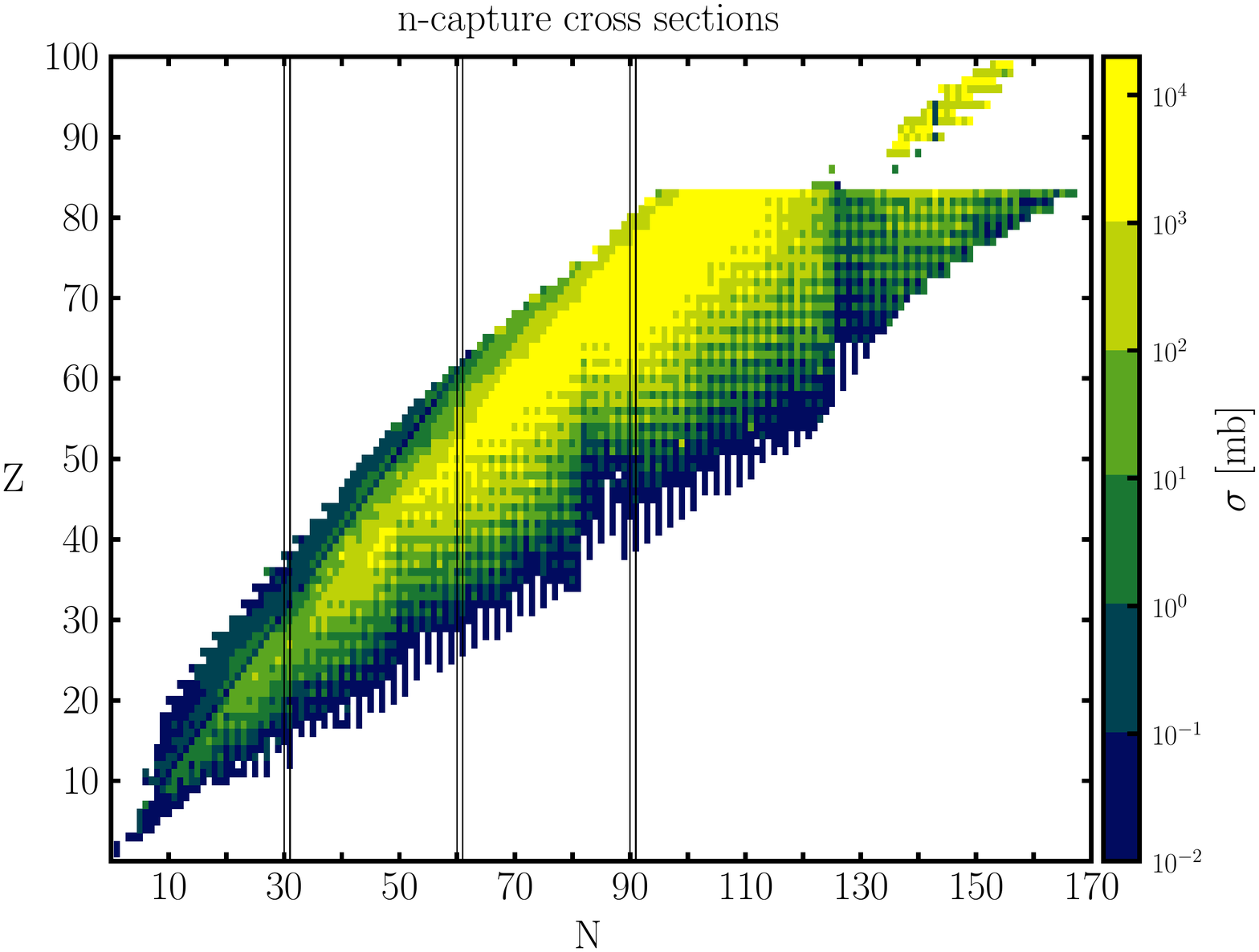}
\includegraphics[width=.49\textwidth]{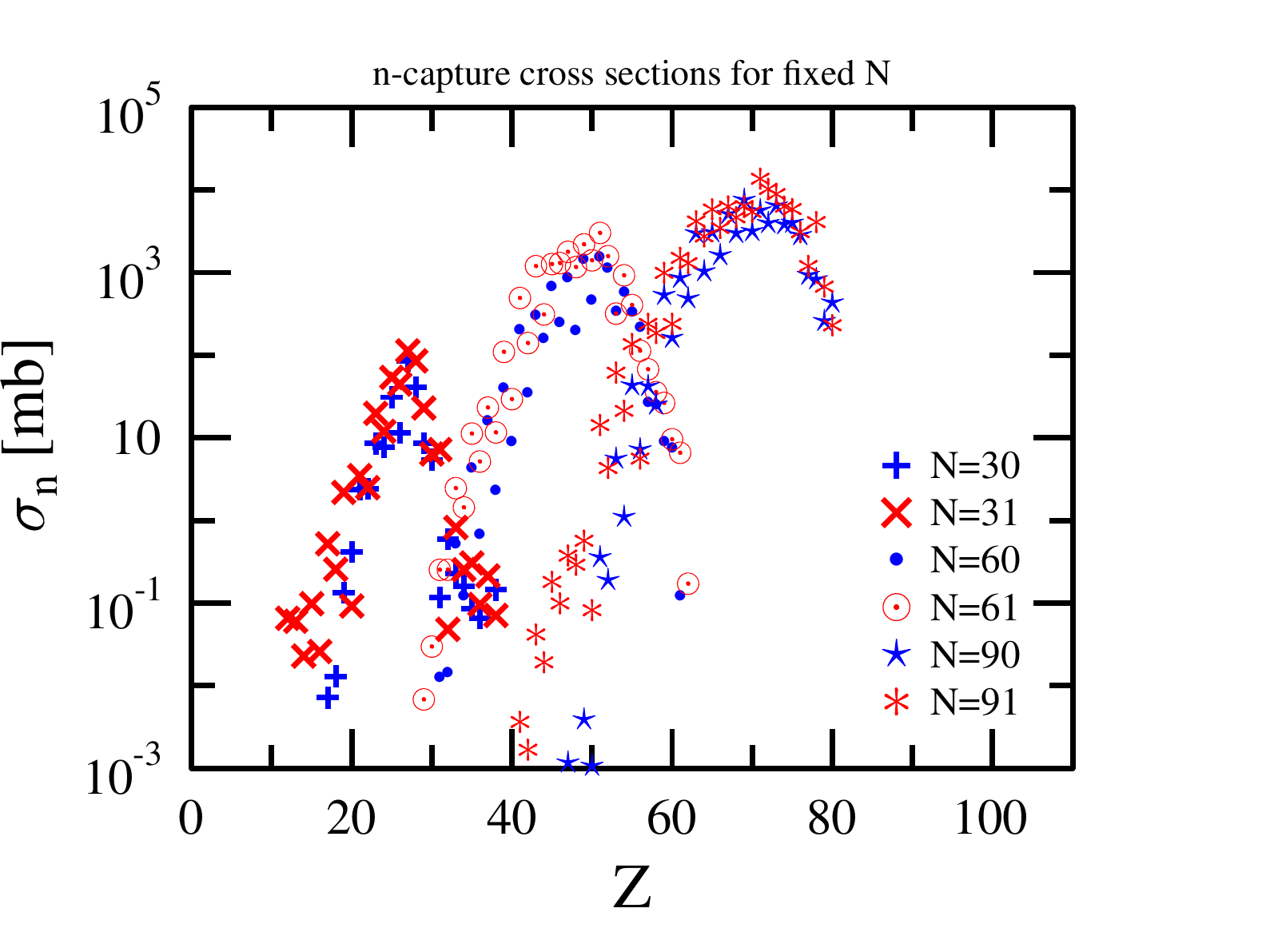}
\caption{(a) MACS (at 30\,keV) on nuclei as a function of the proton
and neutron number.
(b) Dependence on the proton number $Z$ of the MACS on nuclei with
fixed neutron number $N= 30$, 31, 60, 61, 90 and 91 (indicated by
vertical lines on \fig{fig:NCCS30}a).}
\label{fig:NCCS30}
\end{figure}

Maxwellian averaged neutron capture cross sections (MACS) have been
measured for many nuclei and made available in public data
depositories.  A comprehensive and complete review has been presented
recently in \Ref{Pritychenko:2009fe}. Studying the available data, we
can make several observations: (i) although cross sections of many
nuclei have been measured, there are still many missing, or rather
uncertain data, especially for nuclei with $Z > 83$ (see
\fig{fig:NCCS30}a); (ii) the cross sections vary over very large range
of values (about four orders of magnitude); (iii) for any fixed neutron
number $N$ the the cross section is maximal for a corresponding value
of the proton number $\Zmax$ and decreases rapidly as $|Z-\Zmax|$
increases (see \fig{fig:NCCS30}b). The last point implies that in the
$Z-N$ plain for each $N$ their is a unique value $\Zmax(N)$ where the
capture cross section attains its maximal value. The existence of such
a maximum is qualitatively easily understood: for fixed $N$, increasing
$Z$ starting from a small value of $Z$, the capture of an additional
$N$ stabilizes the nucleus in the strong repulsive Coulomb field of the
protons, the binding energy per nucleon increases. However, for $Z$
above some value $\Zmax(N)$ the nucleus developes a neutron skin and
additional neutrons become more and more loosely bound and capturing
any further neutrons becomes less likely. The quantitavie understanding
is certainly more complex, which however, is beyond the scope of the
present paper.

\begin{figure}[ht!]
\includegraphics[width=\textwidth]{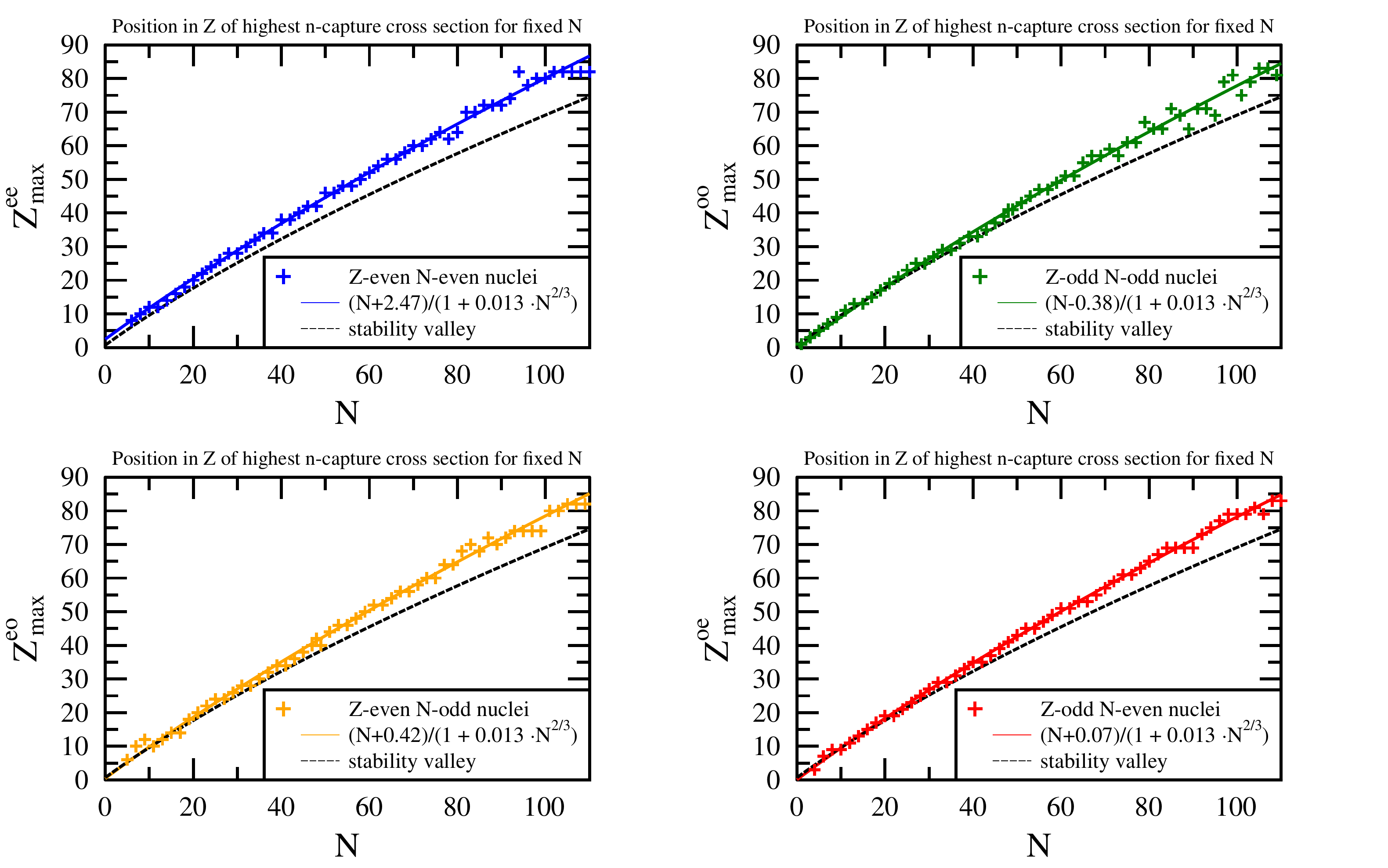}
\caption{The function $\Zmax(N)$ for even-even, odd-odd, even-odd and
odd-even nuclei. The crosses are the experimental values and the solid
line represents the fit to the function in \eqn{eq:f}. The dashed line
runs through the bottom of the stability valley.}
\label{fig:Zmax}
\end{figure}

If we plot the $\Zmax(N)$ function then a rather simple picture
emerges: it appears that a simple, almost linear function can describe
the data, especially for small $N$. This feature becomes even more
salient if we devide the nuclei into four groups according to the
even/odd number of protons and neutrons:
(i) $\Zmax^{({\rm ee})}$ for $Z$ even, $N$ even,
(ii) $\Zmax^{({\rm oo})}$ for $Z$ odd, $N$ odd,
(iii) $\Zmax^{({\rm eo})}$ for $Z$ even, $N$ odd, and 
(iv) $\Zmax^{({\rm oe})}$ for $Z$ odd, $N$ even, as shown in
\fig{fig:Zmax}.

In \fig{fig:Zmax} crosses mark the values of $\Zmax$ where the
n-capture cross section is maximal for a fixed value of the neutron
number $N$ as taken from \Ref{Nakagawa:2005}. The solid lines
represent fits of simple functions to these points in the form of
\beq
f(N;a_x,b,c) = \frac{N+a_x}{1+b N^c}
\,,
\label{eq:f}
\eeq
with $a_x$, $b$ and $c$ being fitted parameters, and $x = $ ee, oo, eo,
or oe. We determined the values of these parameters in two steps. First,
we minimized the function
\beq
\chi^2(a,b,c) =
\sum_{i=1}^{214} \Big(\Zmax(N_i) - f(N;a,b,c)\Big)^2
\,,
\eeq
i.e.\ nuclei belonging to all four groups are taken into account
and all points are assumed to have weight $\sigma_i = 1$. The upper
limit in each group was chosen the largest value for which $\Zmax$
can be identified. With such a choice the we find $N^{(x)}_{\max} = 50$,
56, 53 and 55 maxima in the groups of even-even, odd-odd, even-odd and
odd-even nuclei, respectively ($50 + 56 + 53 + 55 = 214$). This fit
gives
\beq
a = 0.060
\,,\qquad
b = 0.013
\,,\qquad
c = 0.666
\,,
\eeq
with correlation index 
\beq
i = \sqrt{1 - \frac{\chi^2(a,b,c)}
{\sum_{i=1}^{214}\Big(\Zmax(N_i) - \overline{Z}\Big)^2}}
= \sqrt{1 - \frac{590.2}{116029}} = 0.997
\,,
\eeq
i.e.\ the coefficient of determination is almost one, $i^2 = 0.994$
($\overline{Z} = \frac{1}{214} \sum_{i=1}^{214} \Zmax(N_i) = 45.5$).

In the second step we minimize the functions
\beq
\chi^2(a_x) =
\sum_{i=1}^{N^{(x)}_{\max}}
\Big(Z^{(x)}_{\max}(N_i) - f(N;a_x,0.013,0.666)\Big)^2
\eeq
separately for each group ($x = $ ee, oo, eo, oe). These fits result are
\bal
& a_{{\rm ee}} = 2.47
\,,\qquad
a_{{\rm oo}} =-0.38
\,,
& a_{{\rm eo}} = 0.07
\,,\qquad
a_{{\rm oe}} = 0.42 
\,,
\eal
with coefficient of determination above 0.99 in all cases.

We also exhibit the line of the stablity valley in \fig{fig:Zmax},
as a function of $N$ (instead of the usual $A = Z+N$)
\beq
Z_{\rm stab} = \frac{N+a_{\rm s}}{1+b_{\rm s} N^{c_{\rm s}}}
\,,
\eeq
with parameters
\beq
a_{\rm s} = 0.682
\,,\qquad
b_{\rm s} = 0.027
\,,\qquad
c_{\rm s} = 0.614
\,.
\eeq
We see clearly that the highest n-capture cross sections lie above the
stability valley and the separation grows with $N$.

We can also observe regularity in the Z-dependence of the cross section
at fixed N (see \fig{fig:NCCS30}b). We can extrapolate this regularity
as well as the \Zmax\ values to the region in the nuclide chart where
very few data available for n-capture cross sections on nuclei
(nuclei with proton number above 83, see \fig{fig:NCCS30}a.

\begin{figure}[ht!]
\includegraphics[width=\textwidth]{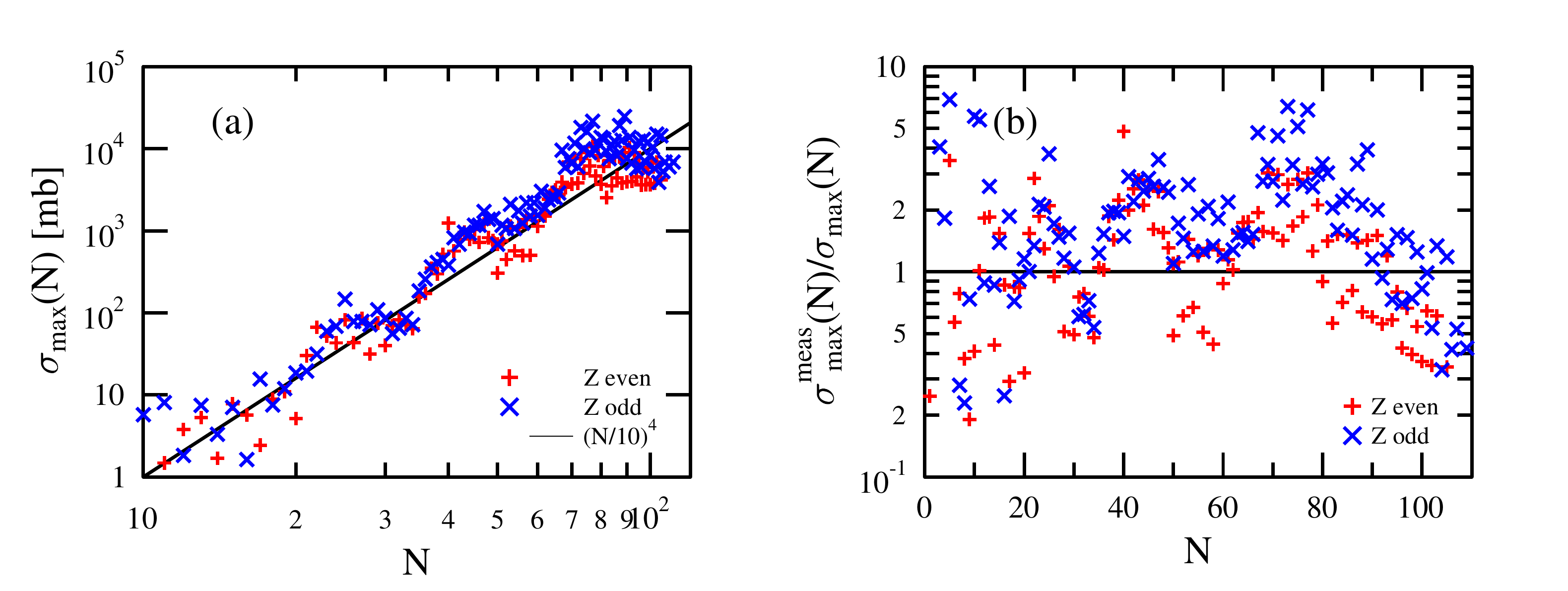}
\caption{(a) Largest neutron capture cross sections as a function of the
neutron number. (b) Ratio of the measured largest cross sections to
$\sigma_{\max}$ given in \eqn{eq:sigmamax}.}
\label{fig:sigmamax}
\end{figure}

The first observation is a simple trend in the behaviour of the
function $\sigma_{\max}(N) \equiv \sigma\Big(\Zmax(N)\Big)$. Putting
$\sigma_{\max}(N)$ on a double logarithmic plot as shown in
\fig{fig:sigmamax}a (left panel), we find that the general trend is well
described by a fourth-order power function,
\beq
\sigma_{\max}(N) = \left(\frac{N}{10}\right)^4\,{\rm mb} \,.
\label{eq:sigmamax}
\eeq
This general trend is slightly modulated with some oscillatory behaviour, 
with minima around magic numbers, as seen on \fig{fig:sigmamax}b, where
the ratios of the measured cross sections to $\sigma_{\max}(N)$ are shown.

\begin{figure}[ht!]
\includegraphics[width=\textwidth]{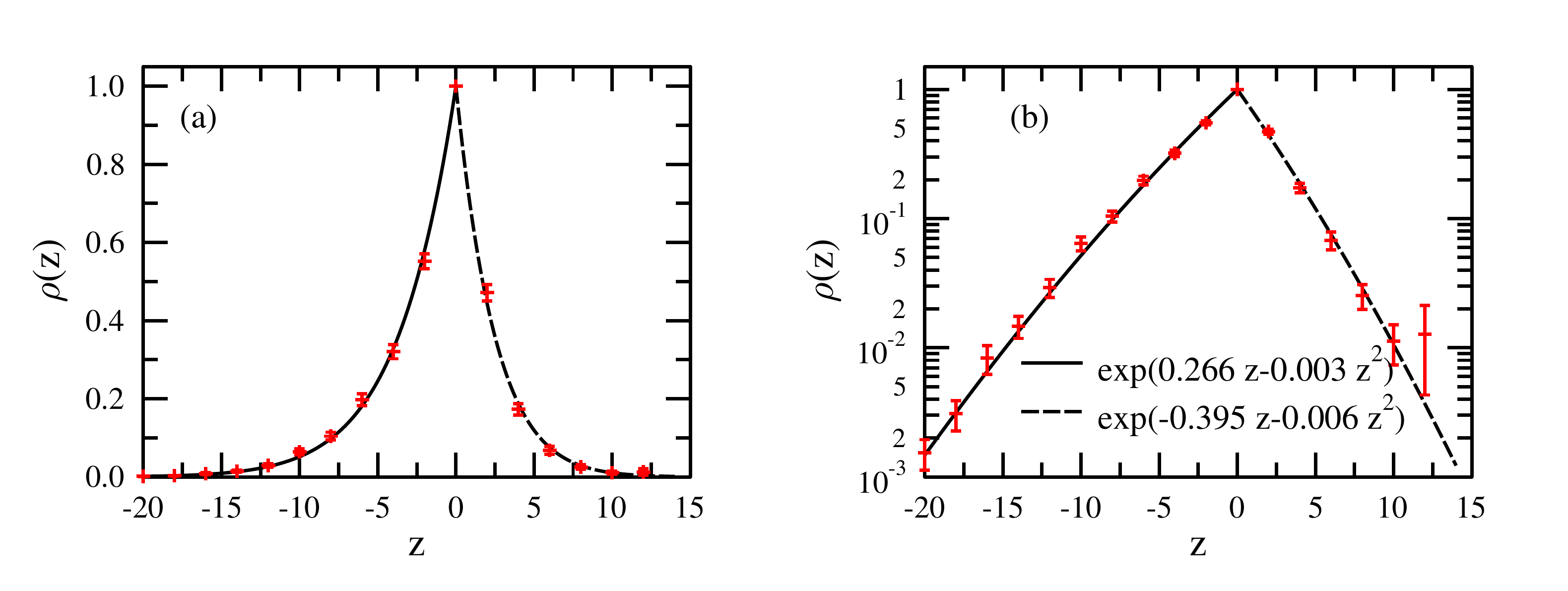}
\caption{Average of the normalized neutron capture cross sections as a
function of $z = Z-\Zmax$. The errorbars represent the standard
deviation $\sigma(z)$.}
\label{fig:rho(z)}
\end{figure}

The second observation is that if we normalize the cross sections
$\sigma(Z,N)$ for a fixed neutron number $N$ with the largest cross
section $\sigma_{\max}(N)$, then the profile of the dependence on the
proton number is rather similar for all neutron numbers. This similarity
is best seen if the position of the largest cross section is shifted by
$-\Zmax$ to zero, therefore, we define these normalized and shifted
cross section values,
\beq
\rho_N(z) = \frac{\sigma(z+\Zmax,N)}{\sigma_{\max}(N)}
\equiv \frac{\sigma(Z,N)}{\sigma\Big(\Zmax(N)\Big)}\,,
\eeq
for all values of $N$, where data are available. Then we define the
average by
\beq
\rho(z) = \frac{1}{N_z} \sum_{N=1}^{N_z} \rho_N(z)\,,
\eeq
with squared standard deviation
\beq
\sigma(z)^2
= \frac{1}{N_z\,(N_z-1)} \sum_{N=1}^{N_z} \Big[\rho_N(z) - \rho(z)\Big]^2
\,,
\eeq
where $N_z$ is the number of available data for fixed $z$.
This average is shown in \fig{fig:rho(z)}. As seen from \fig{fig:rho(z)}b
this function is well approximated with an almost exponential function
in both positive and negative directions, but with different exponents.
More precisely, we fit the logarithm of the average with quadratic
functions of the form $a_i z^2 + b_i z + c_i$ with subscript of the
coefficients refering to three regions in $z$: (i) $i = 1$ for $z<-26$,
(ii) $i = 2$ for $-26\leq z<0$,  and (iii) $i = 3$ for $0<z$. For $i = 2$
and 3 we fix $c_i = 0$. This form ensures the constraint $\rho(0) = 1$.
We also require the continuity of the fitted function at $z = -26$.
We measure the goodness of the fit by the weighted sum of squares
\beq
\chi^2 \simeq
\sum_z \frac{\Big[\ln \rho(z) - (a_i z^2 + b_i z + c_i)\Big]^2}
{\left(\frac{\sigma(z)}{\rho(z)}\right)^2}
\,,
\eeq
summed over values of $z$ in the three regions separately.  The result
of these fits is presented in \tab{tab:rho(z)} and shown in \fig{fig:rho(z)}.

\begin{table}
\caption{\label{tab:rho(z)}
Result of the fit to the average function $\rho(z)$.}
\begin{center}\begin{tabular}{|c|c|c|c|c|}
\hline
\hline
$i$& $a_i$  & $b_i$  & $c_i$ &$\chi^2$/d.o.f  \\ \hline
\hline
1  &  0.0044& 1.135  & 17.95 & 5.29/5 \\ \hline
2  & -0.0025& 0.2658 & 0     & 6.15/9 \\ \hline
3  & -0.0058&-0.3948 & 0     & 7.14/4 \\ \hline
\hline
\end{tabular}
\end{center}
\end{table}

Each function $\rho_N(z)$ differs from the average in two ways: (i)
tipically the larger $N$ the wider $\rho_N(z)$ (as seen on
\fig{fig:NCCS30}a), (ii) in addition there are seemingly random
fluctuations. The origin of the latter could be either a small physical
effect, or simply error of the measurement: there are published values
for cross sections $\sigma(Z,N)$ that differ by a factor of two. While
it is difficult to consider the effect of the latter, the first effect
can be taken into account by a simple appropiate scaling of the width
of the average to those of the functions $\rho_N(z)$, which we discuss
in the next section.

\section{Predictions}

The phenomenological observations made in the previous section can be
used to make predictions for the order of magnitude of neutron capture
cross sections in regions of the nuclide chart where experimental data
are not available. We make these predictions in two steps. First we
validate our procedure by comparing our predictions to measured cross
sections. Then we use our procedure to make predictions.

\subsection{Procedure}

Our procedure relies on three pieces of information concluded from the
analysis of the shape of
{\em ridge of Maxwellian averaged neutron capture cross sections}:
\begin{enumerate}
\itemsep -2pt
\item
position of \Zmax\ as a function of the neutron number (location of the
ridge top on the nuclide chart) obeys the simple function \eqn{eq:f};
\item
values of $\sigma_{\max}(N)$ (height of the ridge for given value of
$\Zmax(N)$) obey the simple function \eqn{eq:sigmamax};
\item
characteristic behaviour of the average function $\rho(z)$ (slope of the
ridge) is as given by \fig{fig:rho(z)}.
\end{enumerate}
In order to predict the cross section values for fixed neutron number, we
proceed along the following steps:
\begin{enumerate}
\itemsep -2pt
\item
Given $N$, find the position of \Zmax\ from \eqn{eq:f}, which gives two
maxima, one for even proton numbers (\Zmaxe) and one for odd proton
numbers (\Zmaxo).  
\item
Given \Zmax\ (either \Zmaxe, or \Zmaxo), position the maximum location
of the average function $\rho(z)$ to \Zmax.
\item
Scale the height and width of the function $\rho(z)$ to the available
measured data by performing a two-parameter fit:
(i) the scale factor of the height, (ii) the scale factor of the width.
\end{enumerate}
The third step is hampered by the discrepancies in the measured cross
section values, which can sometimes be quite significant as shown in
\tab{tab:ratios} for heavy elements. Discrepancies exist among data for
lighter elements, but generally within a factor of two
\cite{Pritychenko:2009fe}.

\begin{table}[t]
\caption{\label{tab:ratios}
Ratios of largest and smallest measured neutron captured cross sections
for elements beyond bismuth \cite{Pritychenko:2009fe}.}
\begin{center}\begin{tabular}{|c|c|c|c|c|c|}
\hline
\hline
nucleus & $\sigma_{\max}/\sigma_{\min}$ &
nucleus & $\sigma_{\max}/\sigma_{\min}$ &
nucleus & $\sigma_{\max}/\sigma_{\min}$ \\
\hline
$^{204}_{82}$Pb$_{122}$ & 1.16 & $^{239}_{92}$U$_{147}$ & 1.42 & $^{245}_{96}$Cm$_{149}$ & 1.19 \\ \hline
$^{207}_{82}$Pb$_{125}$ & 1.25 & $^{240}_{92}$U$_{148}$ & 1.68 & $^{246}_{96}$Cm$_{150}$ & 1.42 \\ \hline
$^{208}_{82}$Pb$_{126}$ & 1.75 & $^{241}_{92}$U$_{149}$ & 2.08 & $^{247}_{96}$Cm$_{151}$ & 1.85 \\ \hline
$^{226}_{89}$Ac$_{137}$ & 1.11 & $^{234}_{93}$Np$_{141}$ & 3.05 & $^{250}_{96}$Cm$_{154}$ & 1.42 \\ \hline
$^{227}_{89}$Ac$_{138}$ & 15.9 & $^{235}_{93}$Np$_{142}$ & 4.09 & $^{245}_{97}$Bk$_{148}$ & 12.4 \\ \hline
$^{228}_{90}$Th$_{138}$ & 1.75 & $^{236}_{93}$Np$_{143}$ & 1.56 & $^{246}_{97}$Bk$_{149}$ & 6.41 \\ \hline
$^{230}_{90}$Th$_{140}$ & 3.32 & $^{238}_{93}$Np$_{145}$ & 16.5 & $^{248}_{97}$Bk$_{151}$ & 7.13 \\ \hline
$^{231}_{90}$Th$_{141}$ & 16.7 & $^{239}_{93}$Np$_{146}$ & 1.55 & $^{250}_{97}$Bk$_{153}$ & 1.72 \\ \hline
$^{234}_{90}$Th$_{144}$ & 2.43 & $^{236}_{94}$Pu$_{142}$ & 3.08 & $^{248}_{98}$Cf$_{150}$ & 3.45 \\ \hline
$^{229}_{91}$Pa$_{138}$ & 4.97 & $^{237}_{94}$Pu$_{143}$ & 1.96 & $^{250}_{98}$Cf$_{152}$ & 1.82 \\ \hline
$^{230}_{91}$Pa$_{139}$ & 1.77 & $^{238}_{94}$Pu$_{144}$ & 1.39 & $^{251}_{98}$Cf$_{153}$ & 1.34 \\ \hline
$^{231}_{91}$Pa$_{140}$ & 2.14 & $^{243}_{94}$Pu$_{149}$ & 1.44 & $^{252}_{98}$Cf$_{154}$ & 2.96 \\ \hline
$^{232}_{91}$Pa$_{141}$ & 2.82 & $^{246}_{94}$Pu$_{152}$ & 12.3 & $^{253}_{98}$Cf$_{155}$ & 20.0 \\ \hline
$^{233}_{91}$Pa$_{142}$ & 1.90 & $^{240}_{95}$Am$_{145}$ & 1.40 & $^{254}_{98}$Cf$_{156}$ & 1.72 \\ \hline
$^{230}_{92}$U$_{138}$ & 4.29 & $^{242}_{95}$Am$_{147}$ & 22.7 & $^{251}_{99}$Es$_{152}$ & 7.48 \\ \hline
$^{231}_{92}$U$_{139}$ & 2.50 & $^{244}_{95}$Am$_{149}$ & 1.34 & $^{252}_{99}$Es$_{153}$ & 2.70 \\ \hline
$^{232}_{92}$U$_{140}$ & 3.38 & $^{242}_{96}$Cm$_{146}$ & 3.33 & $^{253}_{99}$Es$_{154}$ & 46.8 \\ \hline
$^{233}_{92}$U$_{141}$ & 1.40 & $^{243}_{96}$Cm$_{147}$ & 2.03 & $^{254}_{99}$Es$_{155}$ & 5.17 \\ \hline
$^{234}_{92}$U$_{142}$ & 1.27 & $^{244}_{96}$Cm$_{148}$ & 1.55 & $^{255}_{99}$Es$_{156}$ & 3.19 \\ \hline 
$^{237}_{92}$U$_{145}$ & 1.89 & & & & \\ \hline
\hline
\end{tabular}
\end{center}
\end{table}

\subsection{Validation}

\begin{figure}[ht!]
\includegraphics[width=.49\textwidth]{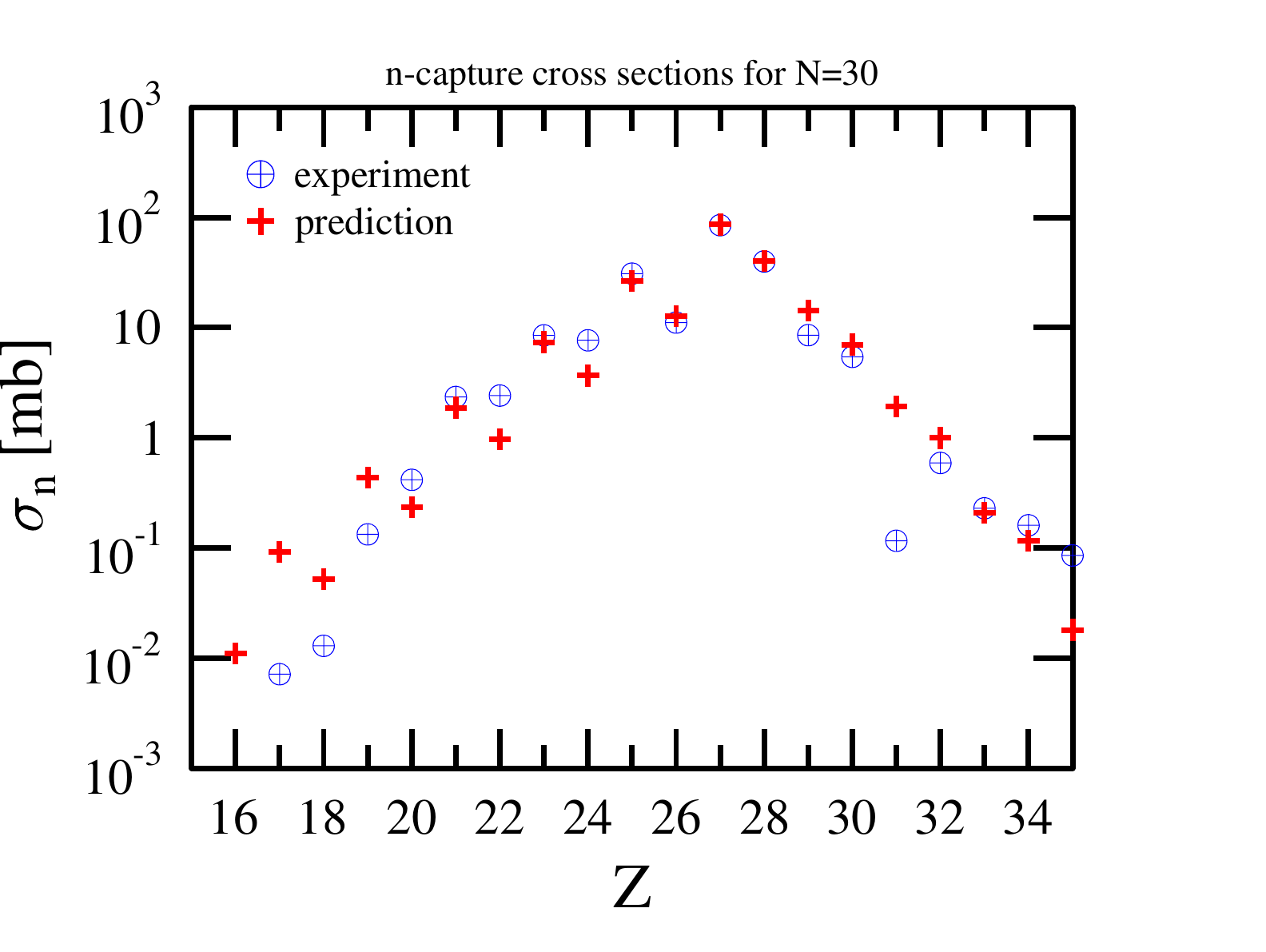}
\includegraphics[width=.49\textwidth]{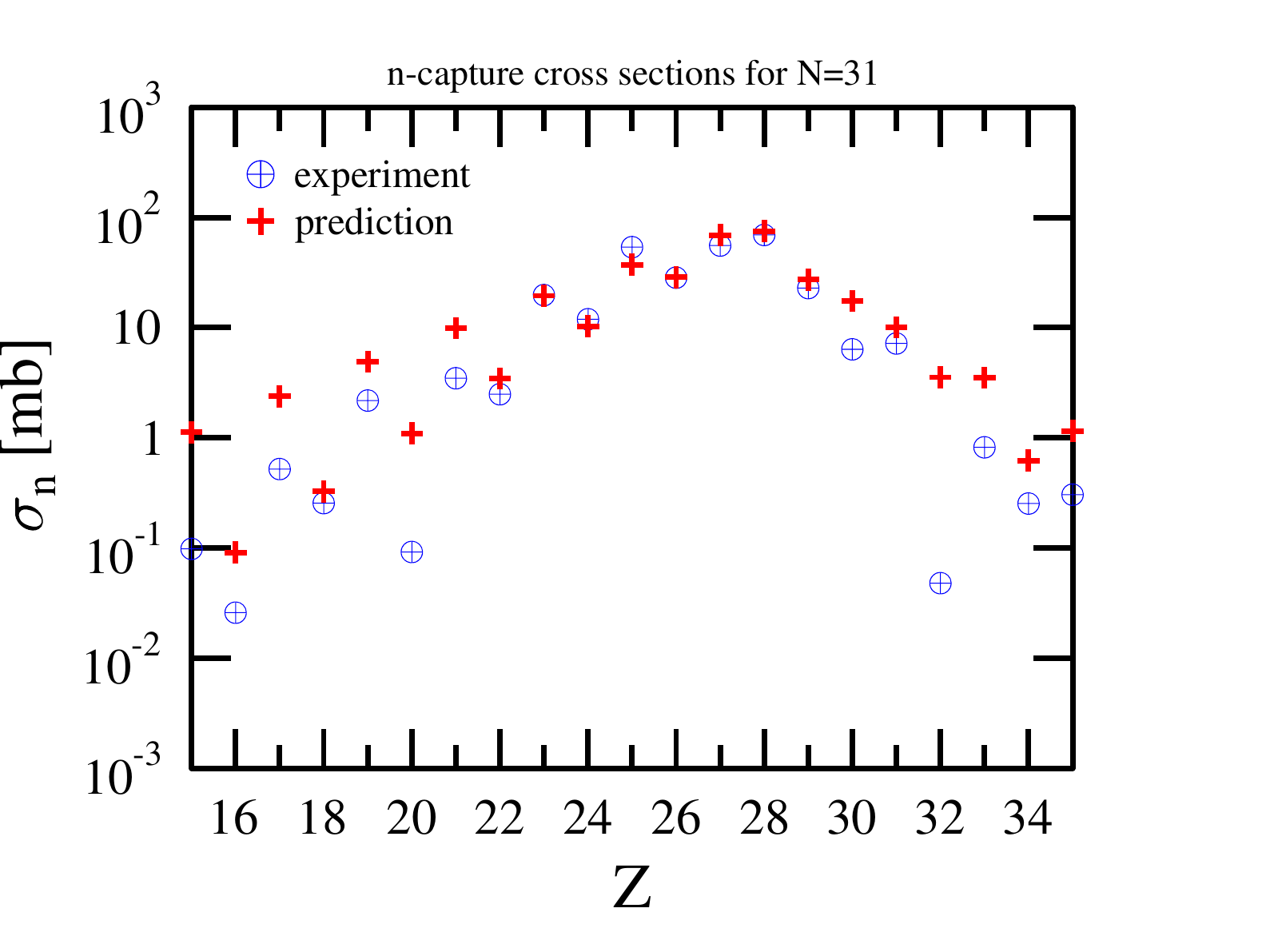}
\includegraphics[width=.49\textwidth]{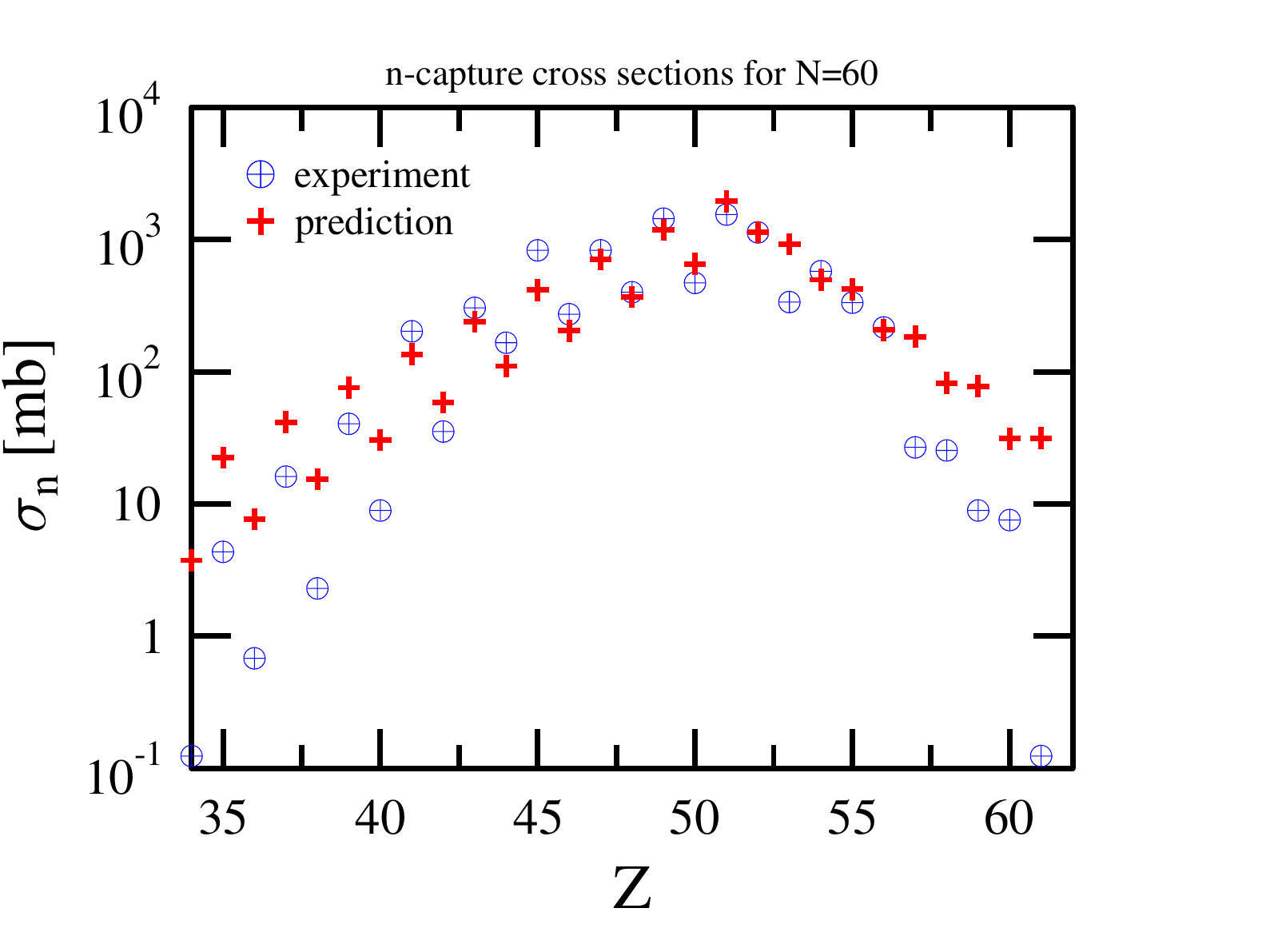}
\includegraphics[width=.49\textwidth]{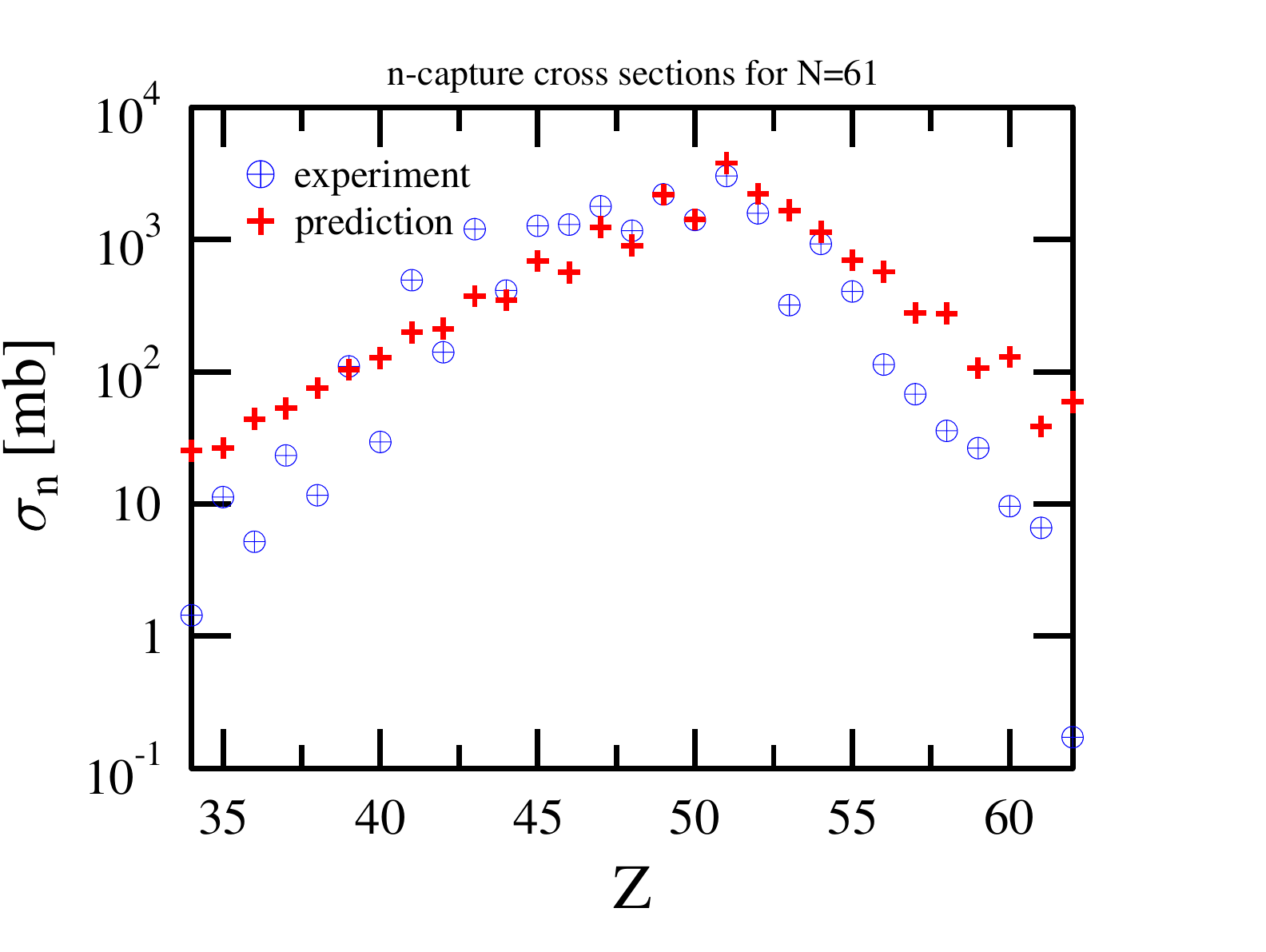}
\includegraphics[width=.49\textwidth]{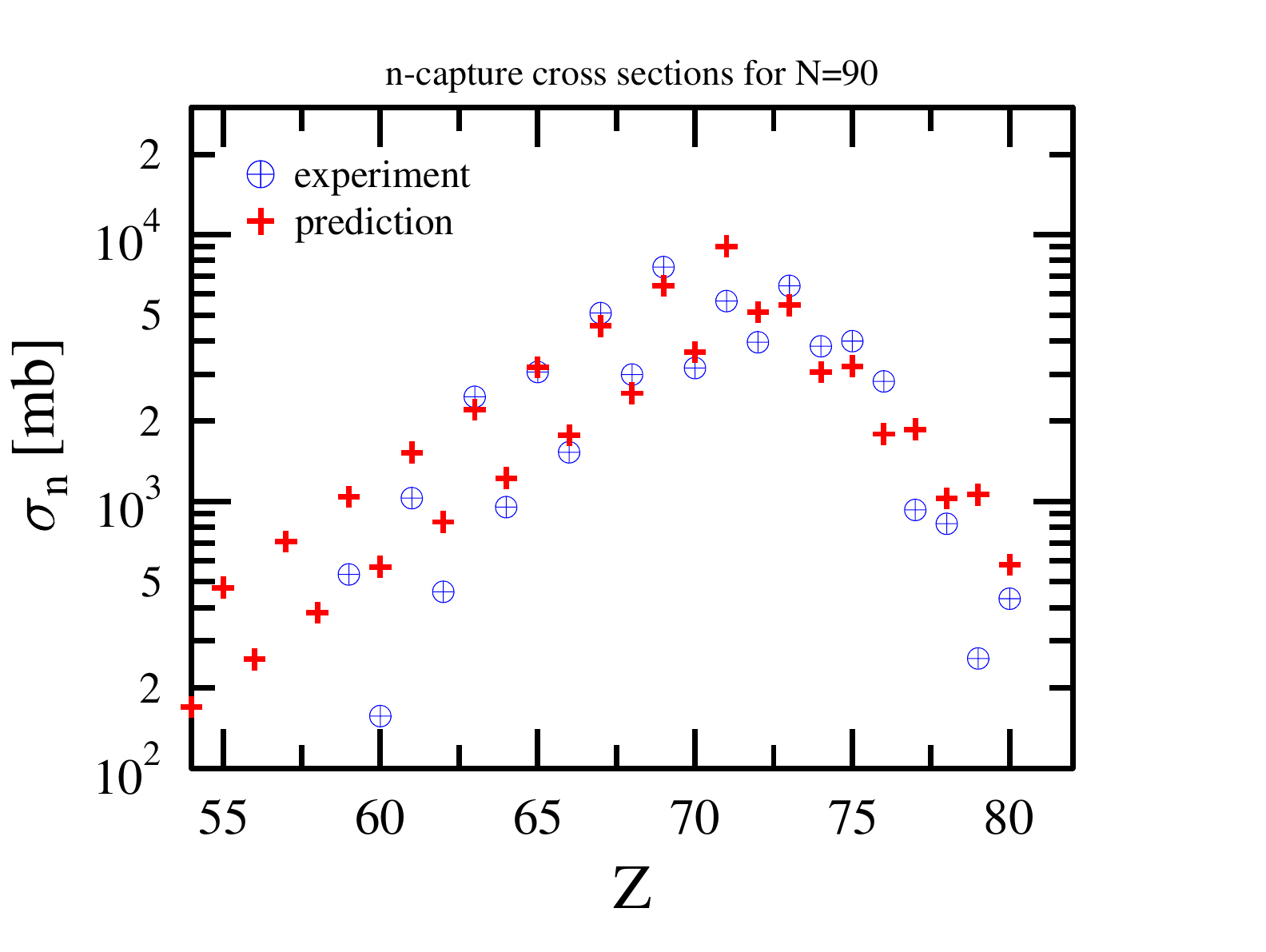}
\includegraphics[width=.49\textwidth]{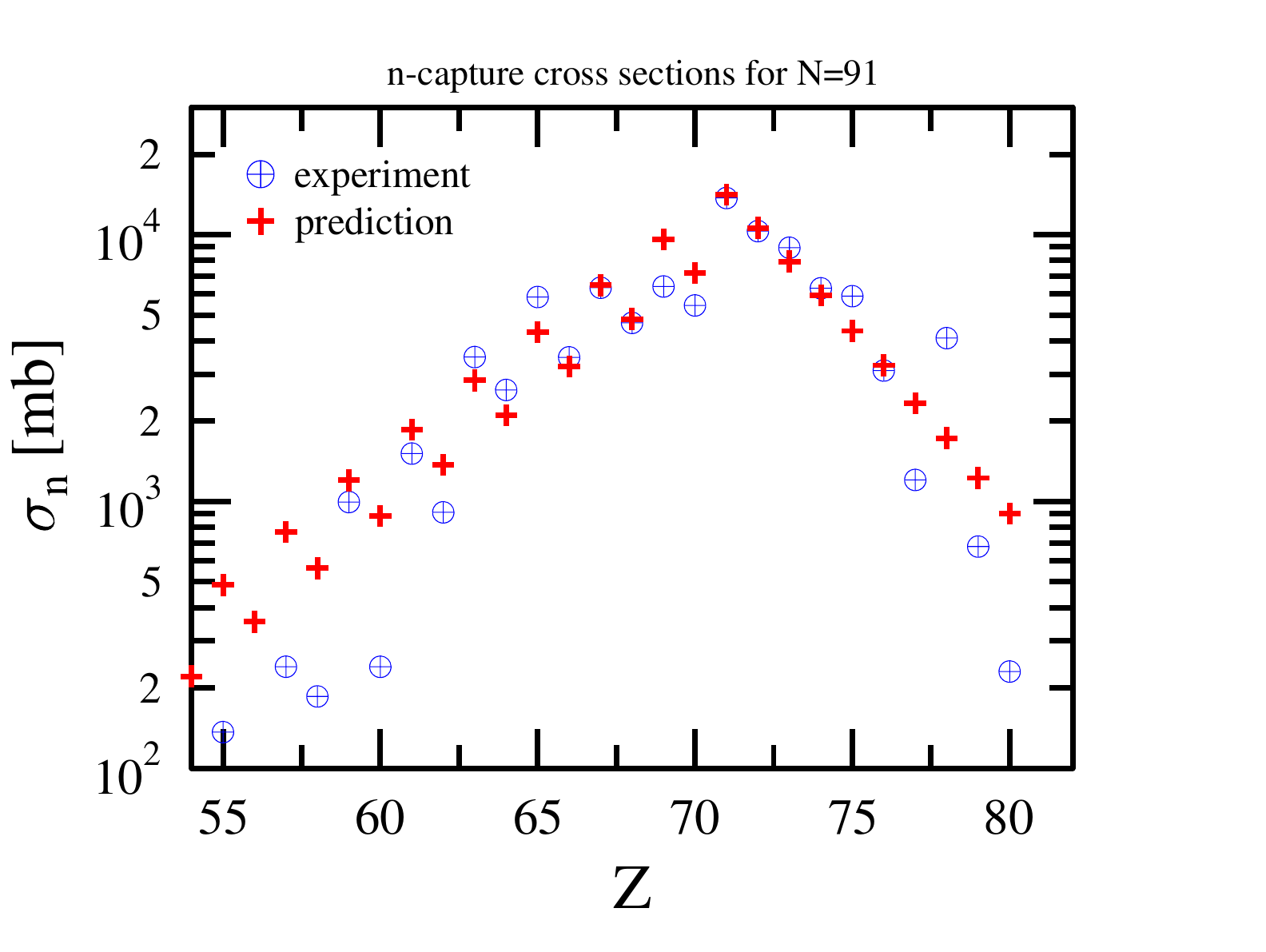}
\caption{
Dependence on the proton number $Z$ of MACS (at 30\,keV) on nuclei with
fixed neutron number $N= 30$, 31, 60, 61, 90 and 91: comparison of the
predictions of the phenomenological model to measure data.
}
\label{fig:N90-91}
\end{figure}

We can compare the values of the predicted cross sections to those
measured experimentally over the regions of the nuclide chart where data
are abundantly available ($Z\le 82$). In \fig{fig:N90-91} we show again
the cross sections of \fig{fig:NCCS30}b together with the predicted
values following from our procedure described in the previous
subsection.  Considering the simple nature of our procedure, the
agreement between data and predictions is striking for all neutron
numbers. Of course, the predictions rarely coincide exactly with
the measurements, but the order of magnitude is usually correct,
especially where the cross sections are large, which is the most
important region for nucleosynthesis. Similar agreement can be
observed over the large region of the nuclide chart where data are
available.  

\clearpage

\subsection{Predictions of unkown cross sections}

\begin{figure}[ht!]
\includegraphics[width=\textwidth]{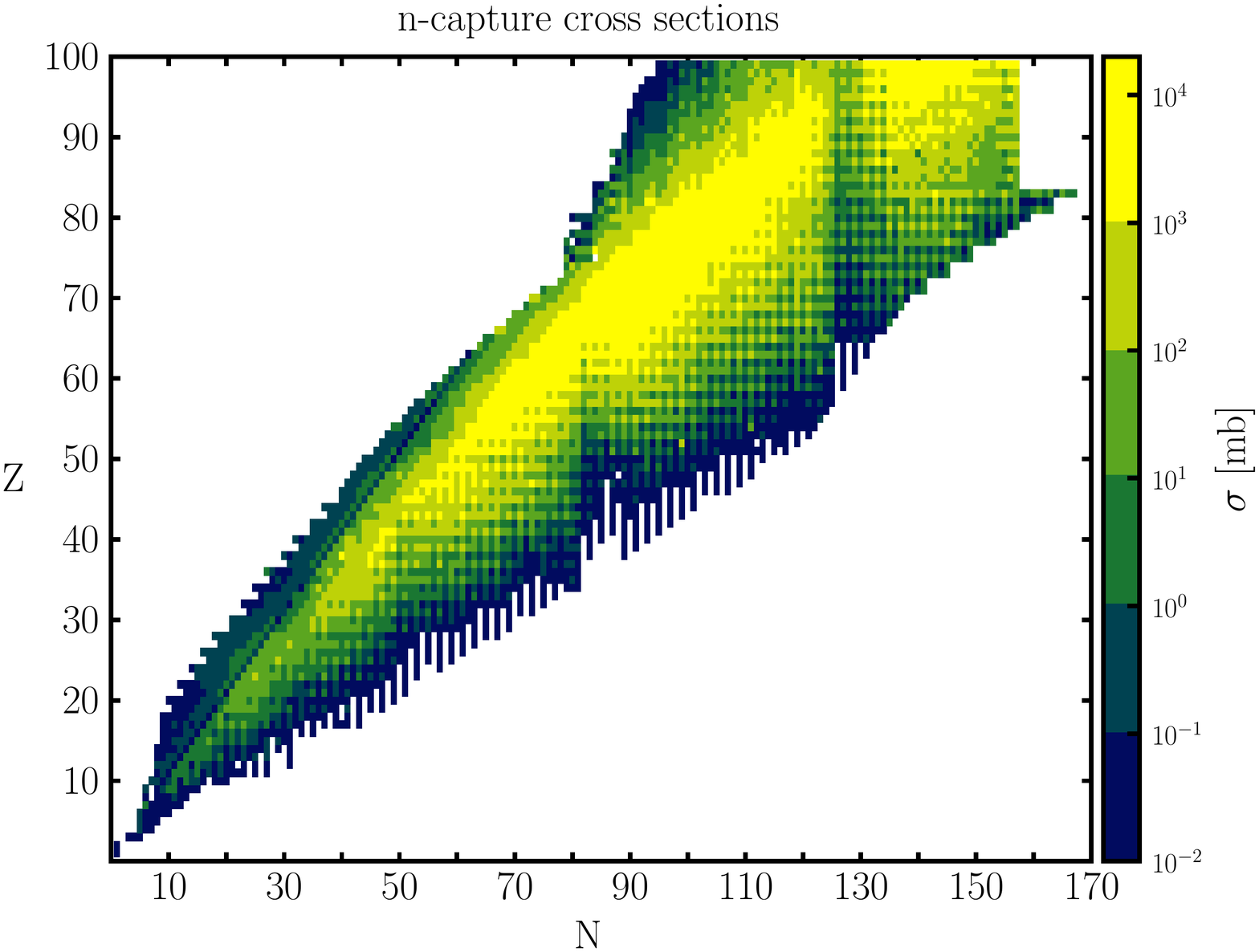}
\caption{Ridge of MACS (at 30\,keV) on nuclei as a function of the
proton and neutron number.}
\label{fig:NCCS30full}
\end{figure}

Our procedure can be used to make prediction for cross sections in
regions of the nuclide chart where {\em some} experimental information are
available, such as $Z>83$. In this region the general trend can be fitted
to the measured data to complete the ridge. With such a procedure we
obtain cross section values shown in \tab{tab:beyondBi}. We can now
use those preditions to complete the picture exhibited on
\fig{fig:NCCS30}. The result of such completion is shown in
\fig{fig:NCCS30full}.

\section{Conclusions}

We studied the dependence of the published MACS data on the proton and
neutron number. We found a simple characteristic behaviour that we call
the shape of the ridge of MACS in the nuclide chart. This shape can be
described by the position and height of the ridge and the decrease of the
slope. Quantifying these characteristics, we made predictions for cross
sections in regions of the nuclide chart where only few data are
available. Such predictions are vital for computer programs aimed at
simulating the formation of heavy elements in stars.

\section*{Acknowledgments}
This research was supported by
the T\'AMOP 4.2.1./B-09/1/KONV-2010-0007 project.
We are grateful to I. Angeli for useful discussions.

\begin{table}
\caption{\label{tab:beyondBi}
Predictions for neutron capture cross sections (in mbarns) as a
function of the proton and neutron number for elements beyond bismuth.}
\begin{center}\begin{tabular}{|r|r|r|r|r|r|r|r|r|r|r|r|}
\hline
\hline
\lower.5ex \hbox{Z}{\Large $\diagdown$}\raise.5ex \hbox{N}
 & 132& 133& 134& 135& 136& 137& 138& 139 & 140& 141& 142\\ \hline
\hline
 84~~~ & 3 & 28 & 3 & 216 & 53 & 331 & 99 & 121 & 90 & 157 & 82 \\ \hline
 85~~~ & 74 & 577 & 99 & 772 & 534 & 229 & 530 & 171 & 385 & 29 & 215 \\ \hline
 86~~~ & 6 & 59 & 5 & 330 & 3 & 512 & 156 & 205 & 150 & 259 & 146 \\ \hline
 87~~~ & 135 & 1024 & 163 & 1276 & 875 & 437 & 792 & 302 & 693 & 74 & 376 \\ \hline
 88~~~ & 13 & 118 & 8 & 601 & 227 & 653 & 448 & 343 & 6 & 421 & 258 \\ \hline
 89~~~ & 242 & 1786 & 266 & 2083 & 1560 & 2020 & 1366 & 523 & 1229 & 184 & 650 \\ \hline
 90~~~ & 26 & 233 & 15 & 751 & 252 & 1400 & 429 & 1400 & 433 & 1550 & 484 \\ \hline
 91~~~ & 428 & 3063 & 428 & 3359 & 2269 & 600 & 1770 & 695 & 2140 & 1213 & 2250 \\ \hline
 92~~~ & 51 & 450 & 25 & 1118 & 412 & 1790 & 427 & 492 & 770 & 425 & 1550 \\ \hline
 93~~~ & 743 & 5168 & 681 & 5347 & 3590 & 2717 & 2514 & 1506 & 3692 & 600 & 1020 \\ \hline
 94~~~ & 98 & 849 & 41 & 1648 & 666 & 2667 & 861 & 1496 & 1036 & 1693 & 750 \\ \hline
 95~~~ & 1267 & 8577 & 1069 & 8406 & 5613 & 4816 & 3635 & 2499 & 6256 & 2345 & 3093 \\ \hline
 96~~~ & 184 & 1568 & 68 & 2407 & 1063 & 3938 & 1289 & 2383 & 1631 & 2635 & 2191 \\ \hline
 97~~~ & 2124 & 14000 & 1657 & 13048 & 8675 & 8379 & 5212 & 4087 & 10442 & 5132 & 5065 \\ \hline
 98~~~ & 336 & 2831 & 111 & 3485 & 1675 & 5761 & 1913 & 3748 & 2538 & 4055 & 3611 \\ \hline
 99~~~ & 3500 & 6718 & 2536 & 20000 & 13250 & 14312 & 7411 & 6584 & 17168 & 10876 & 8185 \\ \hline
100~~~ & 600 & 5000 & 178 & 5000 & 2605 & 8353 & 2813 & 5820 & 3904 & 6172 & 5873 \\ \hline
\hline
\end{tabular}
\end{center}
\end{table}


\begin{thebibliography}{99}

\bibitem{Burbidge:1957vc}
M.~E.~Burbidge, G.~R.~Burbidge, W.~A.~Fowler and F.~Hoyle,
{\em Synthesis of the elements in stars},
Rev.\ Mod.\ Phys.\  {\bf 29}, 547 (1957).

\bibitem{Cameron:1958vx}
A.~G.~W.~Cameron,
{\em Nuclear astrophysics},
Ann.\ Rev.\ Nucl.\ Part.\ Sci.\  {\bf 8}, 299 (1958).

\bibitem{Kappeler:1989}
F.~K\"appeler, H.~Beer and K.~Wisshak
{\em s-process nucleosynthesis -- nuclear physics and the classical model},
Rep.\ Prog.\ Phys.\ {\bf 52}, 945 (1989).

\bibitem{Gallino:1998us}
R.~Gallino {\it et al.},
{\em Evolution and Nucleosynthesis in Low-Mass Asymptotic Giant Branch Stars.
II. Neutron Capture and the s-Process},
Astrophys.\ J.\  {\bf 497}, 388 (1998).

\bibitem{Kappeler:2000}
Z.~Y.~Bao, H.~Beer, F.~K\"appeler, H.~Voss, K.~Wisshak and R.~Tauscher,
{\em Neutron cross sections for nucleosinthesis studies},
Atomic Data and Nuclear Data Tables {\bf 76}, 70 (2000).

\bibitem{kadonis}
I. Dillmann, M. Heil, F. K\"appeler, R. Plag, T. Rauscher and F.-K. Thielemann,
{\em KADoNiS - The Karlsruhe Astrophysical Database of Nucleosynthesis in
Stars},
AIP Conf. Proc. 819, 123; online at http://www.kadonis.org

\bibitem{Nakagawa:2005}
T.~Nakagawa, S.~Chiba, T.~Hayakawa, T.~Kajino,
{\em Maxwellian-averaged neutron-induced reaction cross sections
and astrophysical reaction rates for kT = 1 keV to 1 MeV
calculated from microscopic neutron cross section library JENDL-3.3},
Atom.\ Data Nucl.\ Data Tabl.\  {\bf 91}, 77 (2005).
Available online 28 September 2005.

\bibitem{Pritychenko:2009fe}
B.~Pritychenko, S.~F.~Mughaghab and A.~A.~Sonzogni,
{\em Calculations of Maxwellian-averaged Cross Sections and Astrophysical
Reaction Rates Using the ENDF/B-VII.0, JEFF-3.1, JENDL-3.3 and
ENDF/B-VI.8 Evaluated Nuclear Reaction Data Libraries},
Atom.\ Data Nucl.\ Data Tabl.\  {\bf 96}, 645 (2010)
[arXiv:0905.2086 [astro-ph.SR]].
Available on line at http://www.nndc.bnl.gov/astro/calcmacs.jsp

\bibitem{Rolfs:1988}
C.E.~Rolfs and W.S.~Rodney, 
{\em Cauldrons in the Cosmos},
The University of. Chicago Press, (1988).

\bibitem{Kiss:2010}
M.~Kiss and Z.~Tr\'ocs\'anyi, {\em A unified model for nucleosynthesis of
heavy elements in stars}, J.\ Phys.\ Conf.\ Ser.: Nuclear
Physics in Astrophysics IV, 012024 (2010).
\end{thebibliography}
\end{document}